\documentstyle[12pt]{article}
\def\l#1#2{\raisebox{.2ex}{$\displaystyle
  \mathop{#1}^{{\scriptstyle #2}\rightarrow}$}}
\def\r#1#2{\raisebox{.2ex}{$\displaystyle
 \mathop{#1}^{\leftarrow {\scriptstyle #2}}$}}

\makeatletter
\def\eqnarray{\stepcounter{equation}\let\@currentlabel=\theequation
\global\@eqnswtrue
\global\@eqcnt\z@\tabskip\@centering\let\\=\@eqncr
$$\halign to \displaywidth\bgroup\@eqnsel\hskip\@centering
  $\displaystyle\tabskip\z@{##}$&\global\@eqcnt\@ne
  \hfil$\displaystyle{\hbox{}##\hbox{}}$\hfil
  &\global\@eqcnt\tw@ $\displaystyle\tabskip\z@
  {##}$\hfil\tabskip\@centering&\llap{##}\tabskip\z@\cr}
\@addtoreset{equation}{section}
  \def\theequation{\thesection.\arabic{equation}}
\makeatother

\begin{document}

\renewcommand{\thefootnote}{\fnsymbol{footnote}}
\newpage
\setcounter{page}{0}
\pagestyle{empty}
\begin{flushright}
{November 1996}\\
{JINR E2-96-428}\\
{hep-th/9611148}
\end{flushright}
\vfill

\begin{center}
{\LARGE {\bf The discrete symmetry of the}}\\[0.3cm]
{\LARGE {\bf $N=2$ supersymmetric modified NLS}}\\[0.3cm]
{\LARGE {\bf hierarchy }}\\[1cm]

{\large A. Sorin$^{1}$}
{}~\\
\quad \\
{\em { Bogoliubov Laboratory of Theoretical Physics, JINR,}}\\
{\em 141980 Dubna, Moscow Region, Russia}~\quad\\

\end{center}

\vfill

\centerline{ {\bf Abstract}}
A few new $N=2$ superintegrable mappings in the $(1|2)$ superspace are
proposed and their origin is analyzed. Using one of them, acting like the
discrete symmetry transformation of the $N=2$ supersymmetric modified NLS
hierarchy, the recursion operator and hamiltonian structures of the
hierarchy are constructed.

\vfill
{\em E-Mail:\\
1) sorin@thsun1.jinr.dubna.su }
\newpage
\pagestyle{plain}
\renewcommand{\thefootnote}{\arabic{footnote}}
\setcounter{footnote}{0}

\section{Introduction}

Quite recently the minimal $N=2$ supersymmetric extension of the one
dimensional Toda mapping - f-Toda, has been proposed \cite{dls}.
It acts as the symmetry transformation of the $N=2$
supersymmetric Nonlinear Schr\"odinger (NLS) hierarchy, and almost all
information concerning the hierarchy is encoded in the mapping.

The goal of the present Letter is to continue the list of the $N=2$
superintegrable mappings in the $(1|2)$ superspace in order both to get
new results for the corresponding integrable hierarchies of evolution
equations, and to enlarge the reserve of the mappings for the analysis
of their relations and origin. Besides the connection to integrable
hierarchies, there is one more reason stimulating interest in such
mappings - they are integrable themselves, i.e. every new mapping gives us
a new example of a one-dimensional integrable system.

\section{Note on the $N=2$ supersymmetric mappings}

Before introducing the new mappings, we would like to retrace the origin of
the f-Toda mapping\footnote{The sign $'$ means the
derivative with respect to $x$.}\cite{dls}
\begin{equation}
{1\over 2}(\r {f}{} \r {\overline f}{}-f \overline f) =
(\ln (\overline D \r {f}{}\cdot D
\overline f))'
\label{3}
\end{equation}
for the pair of chiral and
antichiral fermionic superfields $f(x,\theta,\overline \theta)$ and
${\overline f}(x,\theta,\overline \theta)$
\begin{equation}
Df={\overline D}~{\overline f}=0,
\label{2}
\end{equation}
respectively. The notation
$\r {f}{}$ ($\l {f}{}$) means that the index of variable $f$ is shifted by
$+1$ ($-1$) (e.g., $\r {f}{}_n=f_{n+1}$), and we use the standard
representation for the $N=2$ supersymmetric fermionic covariant
derivatives
\begin{eqnarray}
D=\frac{\partial}{\partial \theta}-{1\over 2}\overline \theta \frac{\partial}
{\partial x},\quad \overline D=\frac{\partial}{\partial \overline
\theta}-{1\over 2}\theta\frac{\partial}{\partial x}, \quad
D^2= (\overline D)^2=0, \quad \{D,\overline D \}=
-\frac{\partial}{\partial x} \equiv -{\partial}.
\label{1}
\end{eqnarray}
To do this, let us recall the existence of the mapping \cite{kst} which
connects the $N=2$ super-NLS and the $a=4$ $N=2$ super-KdV
hierarchies \cite{lm}. For example, their second flow equations
\begin{eqnarray}
\frac{\partial}{\partial t_2}f=f''+  D(f{\overline f}~{\overline D}f)',
\quad \frac{\partial}{\partial t_2}{\overline f}=-{\overline f}'' +
{\overline D}(f{\overline f}D{\overline f}),
\label{nls}
\end{eqnarray}
\begin{eqnarray}
\frac{\partial}{\partial t_2}J = [D, {\overline D}] J'+4 J'J,
\label{kdv}
\end{eqnarray}
respectively, are related by the mapping
\begin{eqnarray}
J= -\frac{1}{2} (\frac{1}{2} f {\overline f} + (\ln D {\overline f})').
\label{tr1}
\end{eqnarray}
Equations (\ref{nls}) and (\ref{kdv}) admit the complex structures $*$
\begin{eqnarray}
J^{*}= -J, \quad
{\theta}^{*}={\overline {\theta}}, \quad
{\overline {\theta}}~^{*}={\theta}, \quad
t^{*}_2= -t_2, \quad
x^{*}= x,
\label{conjkdv}
\end{eqnarray}
and
\begin{eqnarray}
f^{*}={\overline f}, \quad
{\overline f}~^{*}=f, \quad
{\theta}^{*}={\overline {\theta}}, \quad
{\overline {\theta}}~^{*}={\theta}, \quad
t^{*}_2= -t_2, \quad
x^{*}= x,
\label{conj}
\end{eqnarray}
respectively, which being applied to the mapping (\ref{tr1}) produce
another admitted mapping
\begin{eqnarray}
J= -\frac{1}{2} (\frac{1}{2} f {\overline f} - (\ln {\overline D} f)').
\label{tr2}
\end{eqnarray}
Denoting the superfields $f$ and $\overline f$ in (\ref{tr2}) by the new
symbols $\r f{}$ and $\r {\overline f}{}$, respectively, and equating the
superfields $J$ belonging to the mappings (\ref{tr1}) and (\ref{tr2}),
we reproduce the f-Toda chain equations (\ref{3}).

The described scheme of getting integrable mappings is quite
general. To illustrate it, let us consider a few new examples.

In \cite{9} the mapping
\begin{eqnarray}
J= \frac{1}{2} (\frac{1}{2} b {\overline b} + (\ln {\overline b})').
\label{mtr1}
\end{eqnarray}
connecting the $N=2$ supersymmetric modified NLS (mNLS) and the $a=4$
$N=2$ super-KdV hierarchies has been constructed. Here,
$b(x,\theta,\overline \theta)$ and
${\overline b}(x,\theta,\overline \theta)$
are chiral and antichiral bosonic superfields
\begin{equation}
Db={\overline D}~{\overline b}=0,
\label{bose2}
\end{equation}
respectively. The second flow equations of $N=2$ super-mNLS hierarhy
\begin{eqnarray}
\frac{\partial}{\partial t_2}b=b''+
D(b{\overline b}~{\overline D}b)', && \quad \frac{\partial}{\partial
t_2}{\overline b}=-{\overline b}'' + {\overline D}(b{\overline
b}D{\overline b})
\label{mnls}
\end{eqnarray}
admit the complex structure
\begin{eqnarray}
b^{*}=i{\overline b}, \quad
{\overline b}~^{*}=i b, \quad
{\theta}^{*}={\overline {\theta}}, \quad
{\overline {\theta}}~^{*}={\theta}, \quad
t^{*}_2= -t_2, \quad
x^{*}= x, \quad
\label{mconj}
\end{eqnarray}
where $i$ is the imaginary unity. Applying (\ref{mconj}) and
(\ref{conjkdv}) to (\ref{mtr1}), we observe that besides the mapping
(\ref{mtr1}) there exists one more
\begin{eqnarray}
J= \frac{1}{2} (\frac{1}{2} b {\overline b} -(\ln b)'),
\label{mtr2}
\end{eqnarray}
and, therefore, the mapping
\begin{equation}
{1\over 2}(\r {b}{} \r {\overline b}{}-b \overline b)=
(\ln (\r {b}{} \overline b))'
\label{bose1}
\end{equation}
acts as the symmetry transformation of the $N=2$ super-mNLS hierarchy.

Quite recently in \cite{dgi} the relationship between
"quasi" $N=4$ SKdV and the $\alpha = -2$, $N=2$ super Boussinesq
\cite{bikp} hierarchies has been established (for detail, see \cite{dgi}).
Using the relations constructed there and following the
above-discussed line, one can produce the mapping
\begin{eqnarray}
&& [D,{\overline D}~]\r V{}+{\r {V}{}}~^{2} +\r {\Phi}{}_+ \r {\Phi}{}_- -
[D,{\overline D}~] V-V^2- {\Phi}_+
{\Phi}_- =-2({\overline D}~\r {V}{}\cdot D+DV\cdot {\overline D})
\ln(\r {\Phi}{}_-{\Phi_+}), \nonumber\\
&& \r {V}{}- V=-(\ln(\r {\Phi}{}_-{\Phi_+}))',
\label{sca}
\end{eqnarray}
which acts as the symmetry transformation of the "quasi"
$N=4$ SKdV hierarchy,
where $V$ is an unconstrained bosonic superfield, and ${\Phi}_+$ and
${\Phi_-}$ are bosonic chiral and antichiral superfields
\begin{equation}
D{\Phi_+}={\overline D}~{\Phi_-}=0,
\label{sca1}
\end{equation}
respectively.

In what follows we restrict ourselves to a concrete example of the mapping
(\ref{bose1}).

The mapping (\ref{bose1}) possesses the inner automorphism
$\sigma$ with the properties
\begin{eqnarray}
\sigma b {\sigma}^{-1}&=& - i \r {\overline b}{}, \quad
\sigma {\overline b} {\sigma}^{-1}= - i \r b{}, \quad
\sigma \r {\overline b}{} {\sigma}^{-1}=i b, \quad
\sigma \r b{} {\sigma}^{-1}= i {\overline b} \nonumber\\
&& \sigma x {\sigma}^{-1}=x, \quad
\sigma {\theta} {\sigma}^{-1}={\overline {\theta}}, \quad
\sigma {\overline {\theta}} {\sigma}^{-1}={\theta}.
\label{au}
\end{eqnarray}

One can rewrite (\ref{bose1}) in a form more similar to (\ref{3})
\begin{eqnarray}
{1\over 2}(D \r {\overline \xi}{} \cdot \overline D \r {\xi}{}-
D \overline \xi \cdot \overline D \xi) =
(\ln (D \r {\overline \xi}{}\cdot \overline D \xi))',
\label{f1}
\end{eqnarray}
if one introduces a new pair of chiral and antichiral fermionic
superfields  $\xi(x,\theta,\overline \theta)$ and
${\overline \xi}(x,\theta,\overline \theta)$
\begin{eqnarray}
D \xi={\overline D}~{\overline \xi}=0,
\label{f2}
\end{eqnarray}
respectively, by the following invertible relations
\begin{eqnarray}
 \xi =& -& {\partial}^{-1} D {\overline b}, \quad
{\overline \xi} =-{\partial}^{-1} {\overline D} b; \nonumber\\
&& b= D {\overline \xi}, \quad {\overline b} = {\overline D} \xi.
\label{f3}
\end{eqnarray}
Relations (\ref{bose1}), (\ref{f1}) and (\ref{f3}) fix only the scaling
dimensions of the products $[b{\overline b}]=cm^{-1}$ and
$[\xi{\overline \xi}]=cm^{0}$. It is interesting to note that despite
the nonlocal character of the mutual relation between the superfields
$b, {\overline b}$ and $\xi, {\overline \xi}$, it will be demonstrated
that the integrable equations of the $N=2$ super-mNLS
hierarchy are local both in terms of the $b, {\overline b}$ and
$\xi, {\overline \xi}$ superfields.

In spite of seeming similarity of the mappings (\ref{bose1}) and
(\ref{f1}) with the f-Toda mapping (\ref{3}), there is an essential
difference between them: neither the mappings (\ref{bose1}) and (\ref{f1})
nor their bosonic limits are algebraically solvable with respect to the
involved superfields $b$, $\overline b$ and $\xi, {\overline \xi}$,
and their bosonic components\footnote{Let us remember that
in the bosonic limit the f-Toda mapping is algebraically solvable.}.

Invariant with respect to the mapping (\ref{bose1}), the hierarchy
of the evolution equations
\begin{equation}
\frac{\partial}{\partial t}
\left(\begin{array}{cc} b \\ {\overline b} \end{array}\right)=
\left(\begin{array}{cc} B\\{\overline B} \end{array}\right),
\label{6}
\end{equation}
with $B$ and ${\overline B}$ being functionals whose independent arguments
are $b,\overline b, b~',{\overline b}~', {\overline D}b,
D{\overline b},...$, can be obtained by solving the symmetry equation,
corresponding to the mapping (\ref{bose1}),
\begin{eqnarray}
{1\over 2}(\r {B}{}_p \r {\overline b}{}+\r {b}{} \r {\overline B}{}_p-
B_p \overline b-b\overline B_p)= ({\r {B}{}_p \over  \r {b}{}} +
{{\overline B}_p \over {\overline b}})'
\label{symbos}
\end{eqnarray}
providing this invariance. In accordance with \cite{5,6}, we got
(\ref{symbos}) taking the derivative of the mapping (\ref{bose1}) with
respect to the time $t$ and substituting eqs.(\ref{6}) into the result.

\section{The list of the results for the mapping (\protect\ref{bose1})}

With short comments, we present the list of the results for the
mapping (\ref{bose1}) in the framework of the general scheme given
in \cite{dls}.

{}~

{\large {\bf 3.1. Solutions of the symmetry equation.}}
The obvious nontrivial solution of the symmetry equation (\ref{symbos}) is
\begin{equation}
\left(\begin{array}{cc} B_0 \\ {\overline B}_0 \end{array}\right)=
\left(\begin{array}{cc} b \\ -{\overline b} \end{array}\right).
\label{solbos}
\end{equation}
The infinite tower of partial solutions $B_p$ and ${\overline B}_p$
($p=0,1,2, ...$) is generated by the recursion
operator $R$
\begin{equation}
\left(\begin{array}{cc}
B_p\\{\overline B}_p
\end{array}\right) = R^{p}
\left(\begin{array}{cc}
B_0 \\ {\overline B_0}
\end{array}\right),
\label{set}
\end{equation}
\begin{eqnarray}
R=\Pi {\partial}\left(\begin{array}{cc}
1- \frac{1}{2}b \partial^{-1} \overline b, & -\frac{1}{2} b \partial^{-1} b
\\-\frac{1}{2} {\overline b}\partial^{-1} \overline b,
& -1-\frac{1}{2}{\overline b}\partial^{-1} b\end{array}\right) \Pi,
\label{recopbos}
\end{eqnarray}
which is defined modulo an arbitrary operator $C{\overline \Pi}$
annihilating the column in the r.h.s. of relation (\ref{set}),
where $C$ is an arbitrary matrix-valued pseudo-differential operator.
Here, $\Pi$ (${\overline {\Pi}}$)
\begin{eqnarray}
\Pi &\equiv& - \left(\begin{array}{cc} D\overline D \partial^{-1}, & 0 \\
0, & \overline D D \partial^{-1}
\end{array}\right), \quad
{\overline {\Pi}} \equiv - \left(\begin{array}{cc} {\overline D}
D \partial^{-1}, & 0 \\
0, & D{\overline D}  \partial^{-1} \end{array}\right), \nonumber\\
&& \Pi \Pi =\Pi, \quad {\overline \Pi}~ {\overline \Pi}=\overline \Pi,
\quad \Pi \overline \Pi=\overline \Pi \Pi=0, \quad \Pi + \overline \Pi = 1
\label{pi}
\end{eqnarray}
is the matrix that projects up and down elements of a column on
the chiral (antichiral) and antichiral (chiral) subspaces, respectively.
To prove (\ref{set}), (\ref{recopbos}) by induction, it is necessary to
use relations (\ref{bose1}) and (\ref{symbos}) as well as their direct
consequences: the two identities which can be obtained from (\ref{bose1})
by the action of derivatives $D$ and $\overline D$, respectively; the
identity which can be produced from (\ref{symbos}) by the action of
$[D,{\overline D}]$; and the following identity
\begin{equation}
({1\over 2} \r {b}{} \r {\overline b}{} -(\ln \r {b}{})')
({1\over 2}{\partial}^{-1}(\r {B}{} \r {\overline b}{}+
\r {b}{} \r {\overline B}{})-
{ \r {B}{} \over{ \r {b}{}}})=
({1\over 2} b{\overline b} + (\ln  {\overline b})')
({1\over 2}{\partial}^{-1}(B{\overline b}+b {\overline B})+
{ {\overline B} \over {\overline b}}),
\label{identbos}
\end{equation}
which one can derive by rewriting relations (\ref{bose1})
and (\ref{symbos}) in the following equivalent form:
\begin{eqnarray}
{1\over 2} \r {b}{} \r {\overline b}{} -(\ln  \r {b}{})'&=&
{1\over 2} b{\overline b} + (\ln  {\overline b})', \nonumber\\
{1\over 2}{\partial}^{-1}(\r {B}{} \r {\overline b}{}+
\r {b}{} \r {\overline B}{})-
{ \r {B}{} \over{ \r {b}{}}}&=&
{1\over 2}{\partial}^{-1}(B{\overline b}+b {\overline B})+
{{\overline B} \over {\overline b}},
\end{eqnarray}
respectively, and equating the product of their left-hand sides to the
product of their right-hand sides.

For example, we present here the first four solutions
\begin{eqnarray}
&& B_0 = b \; , \quad {\overline B}_0 = - {\overline b}; \quad
B_1 = b' \; , \quad {\overline B}_1 = {\overline b}';
\; \nonumber\\ &&B_2 =  b'' +\frac{1}{2}
D{\overline D}(b{\overline b}b), \quad {\overline B}_2 =-{\overline b}'' +
\frac{1}{2}{\overline D}D(b{\overline b}~{\overline b}); \; \nonumber\\
&&B_3 = b''' + D{\overline D}(\frac{3}{2}b'b{\overline b}-
\frac{1}{4}(b{\overline b})^2b), \quad
{\overline B}_3 ={\overline  b}''' -
{\overline D}D(\frac{3}{2}{\overline b}'b{\overline b}+
\frac{1}{4}(b{\overline b})^2 {\overline b}).
\label{threebos}
\end{eqnarray}
These expressions coincide with the corresponding ones
for the $N=2$ super-mNLS hierarchy \cite{9}.

It is instructive to consider the first nontrivial equations (\ref{mnls})
belonging to $N=2$ super-mNLS hierarchy in terms of superfield components
defined as
\begin{eqnarray}
r =  b|, \quad {\overline r}  = {\overline b}|, \quad
{\psi} = {\overline D} b|, \quad
{\overline {\psi}} =  D {\overline b}|,
\end{eqnarray}
where $|$ means the $({\theta}, {\bar\theta})\rightarrow 0 $ limit.
In terms of such components the second flow equations become
\begin{eqnarray}
\frac{\partial}{\partial t_2}r &=& r''-r{\overline r}r'-
r{\psi}{\overline {\psi}}, \quad
\frac{\partial}{\partial t_2}{\psi}=
({\psi}'-r{\overline r}{\psi})', \nonumber\\
\frac{\partial}{\partial t_2}{\overline r} &=&
-{\overline r}''-r{\overline r}~{\overline r}'+
{\overline r}{\psi}{\overline {\psi}}, \quad
\frac{\partial}{\partial t_2}{\overline {\psi}}
=-({\overline {\psi}}'+r{\overline r}{\overline {\psi}})',
\label{comp1}
\end{eqnarray}
and in the bosonic limit, i.e. when the fermionic fields $\psi$ and
$\overline \psi$ are equal to zero, they coincide with mNLS equations of
Ref. \cite{cll}.  After passing to the new fields $g$ and ${\overline g}$
defined by the following invertible nonlocal transformations:
\begin{eqnarray}
g  &=&  r~ {\exp (\frac{1}{2}{\partial^{-1}} (r {\overline r}))}, \quad
{\overline g} = {\overline r}~{ \exp (-\frac{1}{2}{\partial^{-1}}
(r {\overline r}))}; \nonumber\\
r  &=&  g~ {\exp (-\frac{1}{2}{\partial^{-1}} (g {\overline g}))}, \quad
{\overline r} = {\overline g}~{\exp (\frac{1}{2}{\partial^{-1}}
(g {\overline g}))},
\label{comp2}
\end{eqnarray}
eqs.(\ref{comp1}) become
\begin{eqnarray}
\frac{\partial}{\partial t_2}g &=& g''-(g{\overline g}g)'-
g{\psi}{\overline {\psi}}, \quad
\frac{\partial}{\partial t_2}{\psi}=
({\psi}'-g{\overline g}{\psi})', \nonumber\\
\frac{\partial}{\partial t_2}{\overline g} &=&
-{\overline r}''-(g{\overline g}~{\overline g})'+
{\overline g}{\psi}{\overline {\psi}}, \quad
\frac{\partial}{\partial t_2}{\overline {\psi}}
=-({\overline {\psi}}'+g{\overline g}{\overline {\psi}})'.
\label{comp3}
\end{eqnarray}
Simple inspection of eqs.(\ref{comp3}) shows that they are also local, and
their bosonic limit coincides with the derivative NLS (dNLS) equations of
Ref. \cite{kn}. Of course, eqs.(\ref{comp3}) also possess the $N=2$
supersymmetry, however, due to the nonlocal character of the
transformations (\ref{comp2}), it is realized nonlocally.  Thus, in
terms of the fields $g, {\overline g}, \psi$ and ${\overline {\psi}}$, the
hierarchy (\ref{set}), (\ref{recopbos}) can be called the $N=2$
supersymmetric dNLS hierarchy reflecting the name of its first nontrivial
bosonic representative.

{}~

{\large {\bf 3.2. Hamiltonians.}} The infinite set of hamiltonians $H_p$
with the scale dimension $p$ and invariant with respect to the mapping
(\ref{bose1}) can be derived from the formula
\begin{equation}
H_p=\int dxd\theta d{\overline \theta} {\cal H}_p \equiv
\int dxd\theta d{\overline \theta}
{\partial}^{-1}(B_p \overline b+ b{\overline B}_p),
\label{den1}
\end{equation}
which is a direct consequence of the symmetry equation (\ref{symbos})
like its counterpart in the case of the f-Toda mapping (\ref{3})
(for detail, see \cite{dls}). The equation of motion for the first
hamiltonian density ${\cal H}_1$ is also the same
\begin{equation}
\frac{\partial}{\partial t_p} {\cal H}_1 = {\cal H}_p'\; ,
\label{hami0}
\end{equation}
and produces the additional superfield integral of motion
\begin{equation}
{\widetilde H}_1 = \int dx {\cal H}_1.
\end{equation}

Using (\ref{den1}) and (\ref{threebos}) we get, for example, the following
expressions for the first four hamiltonian densities
\begin{eqnarray}
{\cal H}_0 &=& 0, \quad {\cal H}_1 = b {\overline b}, \quad {\cal H}_2 =
2(b'{\overline b}-\frac{1}{2}(b{\overline b})'- \frac{1}{2}(b{\overline
b})^{2}), \nonumber\\ {\cal H}_3 &=& b''{\overline b} +b{\overline b}~''
-b'{\overline b}~'-
\frac{3}{2} b {\overline b}~{\overline D} b\cdot D {\overline b}
+ \frac{3}{2}b{\overline b}(b{\overline b}~'-
b'{\overline b})+\frac{3}{4} (b{\overline b})^{3},
\label{hamibos}
\end{eqnarray}
which coincide with the corresponding hamiltonian densities for the
$N=2$ super-mNLS hierarchy \cite{9}, which again confirms the
above-mentioned relation between the mapping (\ref{bose1}) and $N=2$
super-mNLS hierarchy.

{}~

{\large {\bf 3.3. The mapping (\protect\ref{f1}).}}
Let us very briefly discuss the solutions of the symmetry
equation corresponding to the mapping (\ref{f1}).

Representing the solutions (\ref{set}), (\ref{recopbos}) of the symmetry
equation (\ref{symbos}) in the form of recurrent relations
\begin{eqnarray}
B_{p+1}=D{\overline D}(-B_{p}+b{\cal H}_p), \quad {\overline B}_{p+1}=
{\overline D}D({\overline B}_{p}+ {\overline b}{\cal H}_p), \quad
\left(\begin{array}{cc} B_0 \\ {\overline B}_0 \end{array}\right)=
\left(\begin{array}{cc} b \\ -{\overline b} \end{array}\right)
\label{rel}
\end{eqnarray}
using (\ref{6}) and substituting into (\ref{rel}) the transformation
(\ref{f3}) to the fermionic superfields $\xi, {\overline \xi}$ as well as
introducing the definition of their time derivatives
\begin{equation}
\frac{\partial}{\partial t_p}
\left(\begin{array}{cc} \xi \\ {\overline \xi} \end{array}\right)=
\left(\begin{array}{cc} {\Xi}_p \\ {\overline \Xi}_p \end{array}\right),
\label{f6}
\end{equation}
one can easily find the following recurrent relation:
\begin{eqnarray}
{\Xi}_{p+1}=D({\overline D}{\Xi}_{p}+{\overline D}\xi\cdot{\cal H}_p), \quad
{\overline \Xi}_{p+1}={\overline D}(-D {\overline \Xi}_{p}+
D{\overline \xi}\cdot{\cal H}_p), \quad
\left(\begin{array}{cc} {\Xi}_0 \\ {\overline \Xi}_0 \end{array}\right)=
\left(\begin{array}{cc} -\xi \\ {\overline \xi} \end{array}\right)
\label{relf}
\end{eqnarray}
for the solutions of the symmetry equation corresponding to the mapping
(\ref{f1}). From (\ref{relf}) it is evident that in terms of the
superfields $\xi, {\overline \xi}$ the evolution equations of the $N=2$
super-mNLS hierarchy are also local.

{}~

{\large {\bf 3.4. Hamiltonian structures.}} For the chiral-antichiral
bosonic superfields $b$ and ${\overline b}$, a hamiltonian structure $J$
should be a skew symmetric\footnote{Let us remember the rules of the
adjoint conjugation operation $`T`$: $D^{T}=-D$, ${\overline
D}^{T}=-{\overline D}$, $(MN)^{T}=(-1)^{d_Md_N}N^{T}M^{T}$, where $d_M$
($d_N$) is the Grassman parity of the operator $M$ ($N$), equal to 0 (1)
for bosonic (fermionic) operators. In addition, for matrices it is
necessary to take the operation of the matrix transposition. All other
rules can be derived by using these ones.} $J^{T}=-J$ pseudo-differential
$2$x$2$ matrix operator, which besides the Jacobi identity and the chiral
consistency conditions
\begin{eqnarray}
J \Pi={\overline \Pi} J=0, \quad J {\overline \Pi}= \Pi J=J,
\label{cons}
\end{eqnarray}
should also satisfy the following additional constraint \cite{7}:
\begin{eqnarray}
J(b, {\overline b})={\Phi}J(\r {b}{}, \r {\overline b}{}){\Phi}^T,
\label{inv2}
\end{eqnarray}
which provides the invariance of the hamiltonian equations
\begin{equation}
\frac{\partial}{\partial t}
\left(\begin{array}{cc} b\\{\overline b} \end{array}\right) =
J\left(\begin{array}{cc} {\delta}/{\delta b} \\
{\delta}/{\delta {\overline b}} \end{array}\right) H
\label{hameq}
\end{equation}
with respect to the mapping (\ref{bose1}).
Here, ($\Phi$) ${\hat{\Phi}}$ is the  (invert) matrix of Fr\'echet
derivatives, corresponding to the mapping (\ref{bose1}),\footnote{Here,
the derivatives $\partial$, ${\overline D}$ and $D$ act like operators,
i.e. must be commuted with $b$ and ${\overline b}$.}
\begin{eqnarray}
&&{\Phi}={\phi}_1\otimes {\phi}_2\equiv
\Pi \left(\begin{array}{cc}
1+1/2b L^{-1}{\overline b}
{\partial}^{-1}{\overline D}D \\
-1/2{\overline b} L^{-1}{\overline b}
{\partial}^{-1}{\overline D}D
\end{array}\right){\overline b}~^{-1} \otimes
\left(\begin{array}{cc}
\r {\overline b}{}-2{\partial}\r {b}{}^{-1},
& \r b{}\end{array}\right) \Pi, \nonumber\\
&&{\hat{\Phi}}=A\sigma{\Phi}{\sigma}^{-1}A \equiv
{\hat{\phi}}_2\otimes {\hat{\phi}}_1\equiv
(-A\sigma{\phi}_1 {\sigma}^{-1}) \otimes
(-\sigma {\phi}_2 {\sigma}^{-1}A),
\label{fr}
\end{eqnarray}
where the sign `$\otimes$' stands for the tensor
product and the operator $L$
\begin{eqnarray}
L \equiv {\partial} + \frac{1}{2}b{\overline b} +
\frac{1}{2}{\overline b}{\partial}^{-1}D[{\overline D},b],
\quad [{\overline D}, L]=0
\label{l}
\end{eqnarray}
coincides with the Lax operator of $N=2$ super-mNLS
hierarchy \cite{9}, ${\sigma}$ is the automorphism (\ref{au}), and the
matrix $A$
\begin{equation}
A \equiv \left(\begin{array}{cc} 0, & i \\ i, & 0 \end{array}\right)
\label{a}
\end{equation}
is its matrix of Fr\'echet derivatives. The matrices
${\hat{\phi}}_{1,2}$, ${\phi}_{1,2}$, $\Phi$ and ${\hat\Phi}$ possess the
following properties:
\begin{eqnarray}
&& {\hat{\phi}}_1 {\phi}_1={\phi}_2{\hat{\phi}}_2=1, \quad
\quad {\phi}_1 \otimes {\hat{\phi}}_1=
{\hat{\phi}}_2\otimes{\phi}_2  =
\Phi{\hat\Phi} = {\hat\Phi} \Phi =\Pi, \nonumber\\
&& \Phi \r R{}=R\Phi, \quad {\hat\Phi} R=\r R{} {\hat\Phi}, \quad
\Phi \left(\begin{array}{cc} \r B{} \\
\r {\overline B}{} \end{array}\right)=
\left(\begin{array}{cc}  B \\ {\overline B} \end{array}\right), \quad
{\hat\Phi} \left(\begin{array}{cc} B \\ {\overline B} \end{array}\right)=
\left(\begin{array}{cc} \r B{} \\ \r {\overline B}{} \end{array}\right),
\label{mat2}
\end{eqnarray}
where $B$ and $\overline B$ are arbitrary solutions of the symmetry
equation (\ref{symbos})\footnote{To check relation (\ref{mat2}) for the
first nontrivial solution (\ref{solbos}), it is necessary to remove
ambiguity in the operator ${\partial}^{-1} {\partial} 1$, to appear in
calculations, by setting ${\partial}^{-1} {\partial} 1=({\partial}^{-1}
{\partial}) 1 \equiv 1$.}.

We have a solution of eq.(\ref{inv2}) similar to \cite{dls} for the
first hamiltonian structure
\begin{equation} J_1={\phi}_1\otimes{\partial}{\phi}_1^{T},
\label{hamstr1}
\end{equation}
as well as for the $k$-th hamiltonian structure
\begin{equation}
J_k=R^kJ_1.
\label{hamstrn}
\end{equation}
At $k=2$ eq.(\ref{hamstrn}) reproduces the second hamiltonian structure
$J_2$ \cite{9}
\begin{eqnarray}
J_2=\frac{i}{2}\Pi A {\partial}
\label{j2}
\end{eqnarray}
of $N=2$ super-mNLS hierarchy.

One can check also that expression (\ref{recopbos}) for the recursion
operator can be represented in the following form:
\begin{eqnarray}
R = J_2 J^{*}_1,
\label{recop1}
\end{eqnarray}
where the matrix $J^{*}_1$ is defined as
\begin{eqnarray}
J^{*}_1(b, {\overline b}) =
{{\hat{\phi}}_1}^{T} \otimes {\partial}^{-1} {\hat{\phi}}_1 \equiv
{\overline \Pi} \left(\begin{array}{cc}
{\overline b} \partial^{-1} {\overline b},&
{\overline b} \partial^{-1} b+2\\
b \partial^{-1} {\overline b}-2,&
b \partial^{-1} b
\end{array}\right) \Pi
\label{starj}
\end{eqnarray}
and possesses the following properties:
\begin{equation}
\{J^{*}_1,J_1\}=1, \quad J^{*}_1(\r {b}{}, \r {\overline b}{}) =
A {\sigma}J^{*}_1(b, {\overline b}){\sigma}^{-1} A.
\label{h2}
\end{equation}

Let us note that to get the hamiltonian structures, which are
invariant with respect to the mapping (\ref{f1}), one can apply the
transformation (\ref{f3}) to the hamiltonian structures (\ref{hamstr1}),
(\ref{hamstrn}). Thanks to a very simple structure of its matrix of
Fr\'echet derivatives
\begin{equation}
\left(\begin{array}{cc} 0, &
D \\ {\overline D}, & 0 \end{array}\right)
\label{trfr}
\end{equation}
one can easily find the corresponding formulas, which are not presented
here.

\section{Conclusion}

In this Letter, we proposed a few supersymmetric mappings in the
$(1|2)$ superspace, which act as the symmetry transformation of
the integrable hierarchies corresponding to them, and analyzed their
common origin. Using one of them as an example, we constructed the
recursion operator and hamiltonian structures of the $N=2$ supersymmetric
modified NLS hierarchy. We believe that the same approach can be realized
for other superintegrable mappings and hope to return to it in future.

\section{Acknowledgments}

The author would like to thank Z. Popowicz and especially A.N. Leznov for
many useful and clarifying discussions.

This work was partially supported by grant of the Russian
Foundation for Basic Research RFBR-96-02-17634, INTAS grant
INTAS-94-2317, and by grant from the Dutch NWO organization.

\end{document}